\documentstyle[epsf,12pt,a4]{article}
\input FEYNMAN
\begin{document}
\setcounter{page}{0}
\thispagestyle{empty}
\begin{flushright}
GUTPA-96-4-2
\end{flushright}
\begin{flushright}
NBI-HE-96-15
\end{flushright}
\vskip .2in
\begin{center}
{\large\bf Fermion Masses and Anti-grand Unification\\}
\vskip .4cm
{\bf C. D. Froggatt$^1$, H. B. Nielsen$^2$ and D. J. Smith$^2$}
\vskip .2cm
{ $^1$ \em Department of Physics and Astronomy,
University of Glasgow,\\ Glasgow G12 8QQ, UK}
\vskip .2cm
{$^2$ \em Niels Bohr Institute, Blegdamsvej 17-21,
\\DK 2100 Copenhagen, Denmark}
\end{center}
\section*{ }
\begin{center}
\vskip  .4cm
{\large\bf Abstract}
\end{center}
We present an extension of the Standard Model (SM) without supersymmetry, which
we use to calculate order of magnitude values for the elements of the mass
matrices in the SM. In our model we can fit the 9 quark and lepton masses and 3
mixing angles using only 3 free parameters, with the overall mass scale
set by the electroweak interaction.
The specific model described here has the anti-grand unified
gauge group $SMG^3 \otimes U(1)_f$ at
high energies where $SMG \equiv SU(3) \otimes SU(2) \otimes U(1)$ is the SM
gauge group. The SM fermions are placed in representations of the full gauge
group so that they do not produce any anomalies.
It is pointed out that the same results can be obtained in an
anomaly free
\linebreak $SMG \otimes U(1)^3$ model.
\newpage
\section{Introduction}
\label{intro}

In the Standard Model (SM) all the fermions get a mass via the SM Higgs
mechanism. The Higgs sector of the SM is not satisfactory for two reasons.
First there is a different Yukawa coupling for each fermion and these
couplings make up more than half of the free parameters in the SM. Second, it
would seem natural that these couplings should all be similar, probably of
order 1, since there is no reason in the SM for the
Higgs field to prefer to couple
to one fermion rather than another. However, this is clearly not the case
experimentally where we know there is a huge range of Yukawa couplings from
the top quark with a Yukawa coupling of order 1 to the electron with a
Yukawa coupling of order $10^{-5}$. It is this problem that we wish to address
in this paper and we will introduce an extension of
the SM where such a large range of
Yukawa couplings is natural, due to the existence of new
approximately conserved chiral gauge quantum numbers which protect the fermions
from gaining a mass.
In our model all the elements of the Yukawa
matrices (except for one element which leads to the unsuppressed top mass) are
suppressed (relative to the assumed natural order 1) by a product of the
ratios of the vacuum expectation values (VEVs) of the
Higgs fields required to break the extended gauge symmetry
to the fundamental scale. In this way we can
express very small numbers such as $10^{-5}$ as the product of 5 numbers of
order $10^{-1}$. This type of approach to mass protection
has been considered previously, using abelian chiral charges,
for the SM \cite{fn,bijnens}
and more recently for the minimal supersymmetric SM
\cite{leurer}-\cite{pokorski}.
Here we consider the so-called anti-grand unification model \cite{Bennett}
based on the gauge group $SMG^3 \otimes U(1)_f$, where
$SMG=SU(3) \otimes SU(2) \otimes U(1)$. In fact
the same model was considered in \cite{Gerry2} but in a more abstract manner.
In this paper we further specify the model by actually making a choice of Higgs
particles and we find a much better fit. Some other recent
attempts to model the fermion masses and mixing angles
are reviewed in \cite{corfu}.
\section{A Realistic Model}
\label{model}
\begin{figure}
\begin{picture}(40000,13000)
\THICKLINES

\drawline\fermion[\E\REG](5000,1500)[7000]
\drawarrow[\E\ATBASE](\pmidx,\pmidy)
\global\advance \pmidy by -2000
\put(\pmidx,\pmidy){$b_L$}

\put(12000,0){$\lambda_1$}

\drawline\fermion[\E\REG](12000,1500)[7000]
\drawarrow[\E\ATBASE](\pmidx,\pmidy)
\global\advance \pmidy by -2000
\put(\pmidx,\pmidy){$M_F$}

\put(19000,0){$\lambda_2$}

\drawline\fermion[\E\REG](19000,1500)[7000]
\drawarrow[\E\ATBASE](\pmidx,\pmidy)
\global\advance \pmidy by -2000
\put(\pmidx,\pmidy){$M_F$}

\put(26000,0){$\lambda_3$}

\drawline\fermion[\E\REG](26000,1500)[7000]
\drawarrow[\E\ATBASE](\pmidx,\pmidy)
\global\advance \pmidy by -2000
\put(\pmidx,\pmidy){$b_R$}

\drawline\scalar[\N\REG](12000,1500)[5]
\global\advance \pmidx by 1500
\global\advance \pmidy by 1500
\put(\pmidx,\pmidy){$\phi_{WS}$}
\global\advance \scalarbackx by -530
\global\advance \scalarbacky by -530
\drawline\fermion[\NE\REG](\scalarbackx,\scalarbacky)[1500]
\global\advance \scalarbacky by 1060
\drawline\fermion[\SE\REG](\scalarbackx,\scalarbacky)[1500]

\drawline\scalar[\N\REG](19000,1500)[5]
\global\advance \pmidx by 1500
\global\advance \pmidy by 1500
\put(\pmidx,\pmidy){$W$}
\global\advance \scalarbackx by -530
\global\advance \scalarbacky by -530
\drawline\fermion[\NE\REG](\scalarbackx,\scalarbacky)[1500]
\global\advance \scalarbacky by 1060
\drawline\fermion[\SE\REG](\scalarbackx,\scalarbacky)[1500]

\drawline\scalar[\N\REG](26000,1500)[5]
\global\advance \pmidx by 1500
\global\advance \pmidy by 1500
\put(\pmidx,\pmidy){$T$}
\global\advance \scalarbackx by -530
\global\advance \scalarbacky by -530
\drawline\fermion[\NE\REG](\scalarbackx,\scalarbacky)[1500]
\global\advance \scalarbacky by 1060
\drawline\fermion[\SE\REG](\scalarbackx,\scalarbacky)[1500]

\end{picture}
\vskip .3cm
\caption{Feynman diagram for bottom quark mass in the full theory.
The crosses indicate the couplings of the Higgs fields to the vacuum.}
\label{MbFull}
\end{figure}
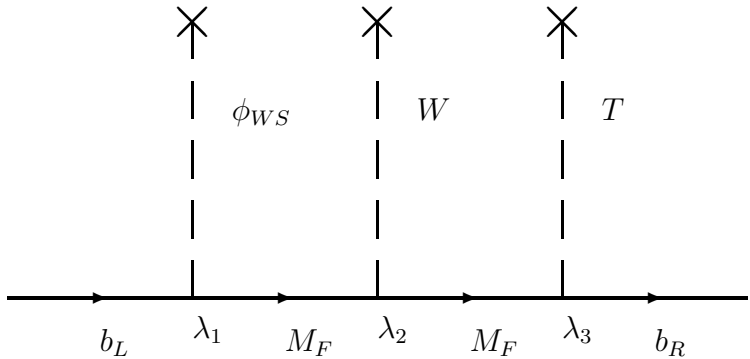

A natural way to explain why there are small Yukawa couplings in the SM is to
assume that the SM is just a low energy effective theory. Then the smallness
of some SM Yukawa couplings can be explained by suppressing the corresponding
interactions in the full theory. A simple way to do this is to extend the SM
gauge group and include the SM fermions in this group so that the quantum
number differences between the left-handed
and right-handed fermions are not the
same as they are in the SM (in the sense that we introduce new quantum numbers
in the full gauge group but, of course, the SM quantum numbers which will be
produced by particular combinations of these quantum numbers will be the usual
SM quantum numbers for all the fermions). In this way we can
still arrange that the
left-handed and right-handed top quarks should differ by
just the quantum numbers of the
Weinberg-Salam Higgs field, and so the top quark mass will
not be suppressed. However, in a
realistic model where all the fundamental Yukawa couplings are of order 1,
no other combination of left-handed and right-handed fermions should
differ by these quantum numbers, as no other fermion has the same order of
magnitude mass as the top quark.
As long as the new chiral gauge quantum numbers are conserved, the
other fermion mass terms are forbidden.
For these other fermion masses we have to
introduce the new Higgs fields, which spontaneously break
the extended gauge symmetry and allow the transitions giving mass terms
in the SM. For example, consider the bottom quark mass.
In the SM this
requires a bottom quark Yukawa coupling constant
significantly less than 1. However, if we
consider that in the full theory the transition amplitude
between the left-handed and right-handed bottom quarks is given
by the Feynman diagram in fig.~\ref{MbFull},
we no longer need Yukawa couplings less than 1.
If we assume that all the {\em fundamental} Yukawa
coupling constants are of order unity, $\lambda_i \sim 1$,
this diagram gives the following
relation for the {\em effective} SM bottom Yukawa coupling constant:
\begin{equation}
h_b \approx \frac{<W>}{M_{F}}
\frac{<T>}{M_{F}}
\label{byukawa}
\end{equation}
where $<W>$ and $<T>$ are simply the VEVs of the new
Higgs fields $W$ and $T$, and
$M_F$ is the (fundamental) mass scale of the intermediate fermions.
Henceforth we will consider the VEVs of all the new Higgs fields
measured in units of $M_F$. The results for the fermion masses are
rather insensitive to the precise value of $M_F$ but, in the
spirit of the anti-grand unification model \cite{Bennett},
we shall assume that $M_F$ is
approximately equal to the Planck mass. This means that the
above relation, eq. (\ref{byukawa}), will be
taken to hold at the Planck scale.

For the smallest masses we shall require several such Higgs fields to be used
rather than just one or two with a very small VEV.
To calculate the effective Yukawa
coupling in the SM, we simply multiply the VEVs of all the new
Higgs fields needed to generate
the interaction in the full theory. By using different combinations
of a few Higgs fields we can reduce the number of free parameters.
So, in a realistic model, the range of masses would be explained
using a few Higgs fields with VEVs within an order
of magnitude of the fundamental scale $M_F$.

To make such a model, we must first choose a gauge group which contains the SM
gauge group (SMG). Then we must put the SM fermions in the group so that they
no longer have the same quantum number differences between left and right
components for
different generations. Finally,
we need to choose some Higgs fields to break the group down to the SMG and also
give realistic masses to the fermions, assuming that all
the fundamental Yukawa couplings are
of order 1. We also need suitable intermediate fermions for diagrams such as
fig.~\ref{MbFull}. These should be vector-like
Dirac fermions with a mass of order
$M_F$. We shall assume that all such fermions required are actually
present; assuming some not to be present would cause extra suppression of some
transitions.
\section{The Anti-grand Unification Model}
\label{SMG3U1}
The anti-grand unification model has previously been considered as a
candidate for explaining the fermion masses \cite{Gerry2}. The gauge group
for the model is:
\begin{equation}
G=SMG_1 \otimes SMG_2 \otimes SMG_3 \otimes U(1)_f
\end{equation}
where we have defined:
\begin{equation}
SMG_i=SU(3)_i \otimes SU(2)_i \otimes U(1)_i
\end{equation}
The three $SMG_i$ groups will be broken down to their diagonal
subgroup, which is
the gauge group of the SM. The $U(1)_f$ group will be totally broken. This
gauge group (strictly speaking without the $U(1)_f$ group) has been used to
successfully predict the values of the running gauge coupling constants
in the SM at the Planck scale\cite{Bennett}, as critical couplings estimated
using lattice gauge theory.

We put the SM fermions into this group in an obvious way. We have one
generation of fermions coupling to each $SMG_i$ in exactly the same way as they
would couple to the SMG in the SM. The broken chiral gauge quantum numbers
of the quarks and leptons under the symmetry groups $SMG_i$ can readily
explain the mass differences between fermion generations but
cannot explain all the mass splittings within each generation,
such as the ratio of the top and bottom quark masses. It is for this
reason that the abelian flavour group $U(1)_f$ is introduced.
We then choose $U(1)_f$ charges with the constraint that there should
be no anomalies and no new mass-protected fermions.
This leads us almost uniquely to
the set of charges shown in table~\ref{Q_f}. We have labelled the fermions
coupling to $SMG_i$ by the names of the `i'th generation of SM fermions.
However, this is just a method of labelling the representations of the full
gauge group and as we will show later, for example,
the fermion we have labelled
$c_R$ will in fact turn out to be the right-handed top quark in the SM.

\begin{table}
\caption{$U(1)_f$ charges of the fermions.}
\begin{displaymath}
\begin{array}{|c|c|c|c|c|c|}
\hline
{\rm Fermion} & Q_f & {\rm Fermion} & Q_f & {\rm Fermion} & Q_f \\ \hline
u_L & 0 & c_L & 0 & t_L & 0 \\ \hline
u_R & 0 & c_R & 1 & t_R & -1 \\ \hline
d_R & 0 & s_R & -1 & b_R & 1 \\ \hline
e_L & 0 & \mu_L & 0 & \tau_L & 0 \\ \hline
e_R & 0 & \mu_R & -1 & \tau_R & 1 \\ \hline
\end{array}
\end{displaymath}
\label{Q_f}
\end{table}
Now we must choose appropriate Higgs fields to break $G$ down to the SMG.
The quantum numbers of the fermion fields are determined by the
theoretical structure of the model (in particular the requirement of
anomaly cancellation), but we do have some freedom in the choice
of the quantum numbers of the Higgs fields.
\subsection{Choosing Higgs Fields}
\label{choosinghiggs}
There are obviously many different ways to break down the large group $G$ to
the much smaller SMG. In order to decide how we should do this,
we must consider
how we are going to suppress the fermion masses in order to get a realistic
model. However, we can first greatly simplify the situation by considering only
the $U(1)$ charges.

To give some justification for considering only the abelian charges,
we first make the observation that in the
SM there is a charge quantisation rule:
\begin{equation}
\frac{y}{2}+\frac{1}{2}{\rm ``duality"}+\frac{1}{3}{\rm ``triality"}
	\equiv 0 \pmod 1
\label{SMGiChQu}
\end{equation}
where $y$ is the conventional weak hypercharge.
If this is not an accident then we must assume that some such quantisation
should be present in the full theory. The simplest way to ensure
this charge quantisation rule, eq.~(\ref{SMGiChQu}), is to
assume that such a rule holds for each $SMG_i$ separately.

The important point now is that if we assume that the
matter (fermion and scalar) fields belong to singlet or
fundamental representations of all non-abelian gauge groups, we can use the
charge quantisation rules to determine the non-abelian representations from
the $U(1)_i$ charges $y_i$. So now the four abelian charges can be used to
specify the complete representation of $G$.
The charges of the Higgs fields are selected
by examining the abelian charge differences between left-handed
and right-handed fermions and choosing combinations of
charges to allow such transitions between them. The
constraint that we must eventually recover the SMG as the diagonal subgroup of
the $SMG_i$ groups is equivalent to the constraint that all the Higgs fields
(except for the Weinberg-Salam Higgs field
which of course finally breaks the SMG)
should have charges $y_i$ satisfying:
\begin{equation}
\label{diagU1}
y=y_1+y_2+y_3=0
\end{equation}
in order that their SM weak hypercharge $y$ be zero.
\subsection{The Weinberg-Salam Higgs Field}
\label{WSHiggs}
We wish to choose the charges of the Weinberg-Salam (WS) Higgs
field so that it matches the difference in charges between
the left-handed and right-handed top
quarks. This will ensure that the top quark
mass in the SM is not suppressed relative
to the WS Higgs field VEV. However, there is
a problem with this. If we choose the left-handed
and right-handed top quarks to be the fermions previously labelled
$t_L$ and $t_R$ then we cannot suppress the bottom and tau masses. This is
because the charge differences between $t_L$ and $t_R$
are the same as between $b_L$ and $b_R$ and also
between $\tau_L$ and $\tau_R$.

The only solution to this problem in our model is to choose the left-handed and
right-handed top quarks to be Weyl fermions from
different ``proto-generations'', i.~e.\ one but not both Weyl states
should couple to $SMG_3$. In particular, we can choose
the right-handed top quark to be any of the 3 fermions we have labelled $u_R$,
$c_R$ and $t_R$, since these fermions will
have exactly the same quantum numbers
in the SM. So we will choose the top mass to be due to the transition between
the fermions we have labelled $t_L$ and $c_R$.

Now it is simple to calculate the quantum numbers of the WS Higgs field.
We will use a notation to label the 4 abelian charges by:
\begin{displaymath}
\vec{Q} = \left( \frac{y_1}{2}, \frac{y_2}{2}, \frac{y_3}{2}, Q_f \right)
\end{displaymath}
where $\frac{y_i}{2}$ are the weak hypercharges with respect to the groups
$U(1)_i$ (normalised as in the usual SM convention) and $Q_f$ is the charge
with respect to the group $U(1)_f$. So the quantum numbers of the WS Higgs
field are:
\begin{equation}
\vec{Q}_{WS} = \vec{Q}_{c_R} - \vec{Q}_{t_L}
	= \left( 0,\frac{2}{3},0,1 \right) - \left( 0,0,\frac{1}{6},0 \right)
	= \left( 0,\frac{2}{3},-\frac{1}{6},1 \right)
\end{equation}
This means that the WS Higgs field
will in fact be coloured under both $SU(3)_2$ and
$SU(3)_3$. After breaking the symmetry down to the SMG,
we will be left with the usual WS Higgs field of the SM and another scalar
which will be an octet of $SU(3)$
and a doublet of $SU(2)$.
This should not present any phenomenological problems,
provided this scalar doesn't cause symmetry breaking and doesn't have a mass
less than the electroweak scale. In particular an octet of $SU(3)$
cannot lead to baryon decay.
\subsection{Choosing the Other Higgs Fields}
\label{OtherHiggs fields}
We can now choose the charges of the other Higgs fields in our
model, by considering
the charge differences between left-handed and right-handed fermions with the
inclusion of the WS Higgs. Since we have the constraint of eq.~(\ref{diagU1}),
the charges of these Higgs fields must be chosen to span a 3 dimensional vector
space of charges represented, for example,
by $\frac{y_1}{2}$, $\frac{y_3}{2}$ and
$Q_f$ with $\frac{y_2}{2}$ being determined by eq.~(\ref{diagU1}). (These
charges are quantised in units of $\frac{1}{6}$ for $\frac{y_1}{2}$ and
$\frac{y_3}{2}$ due to eq.~(\ref{SMGiChQu}) and the $Q_f$ charges
are normalised to be integers for fermions as in
table~\ref{Q_f}). This means that we will need at
least 3 Higgs fields to break the gauge
group down to the SMG. This gives us a lot of
freedom, so we will choose the charges on these Higgs fields by considering
phenomenological relations between fermion masses.

Since we are assuming that the fundamental Yukawa couplings are of order 1 but
not exactly 1, we can only produce order of magnitude results. So we wish to
choose, for example, 2 fermions with similar masses
but not order of magnitude equal masses.
We can then assume that the lighter fermion is suppressed relative to the
heavier fermion by 1 Higgs with a VEV given
approximately by the ratio of the 2 fermion masses. For example we would say
that the bottom quark and tau lepton masses were of the
same order of magnitude (remembering
that we take all relations at the Planck scale). However we can take the
following 2 ratios of effective Yukawa couplings to be significantly different
from 1:
\begin{equation}
\frac{h_c}{h_b} \equiv <W> \approx \frac{1}{5}
\label{mc/mb}
\end{equation}
\begin{equation}
\frac{h_{\mu}}{h_b} \equiv <T> \approx \frac{1}{13}
\label{mmu/mb}
\end{equation}
where we have defined 2 Higgs fields, $W$ and $T$, to have appropriate VEVs to
cause the charm and muon to be suppressed relative to the bottom.

First we define $\vec{b}$ to be the difference in charges
between $b_L$ and $b_R$ with the inclusion of the WS Higgs field. So we have:
\begin{equation}
\vec{b} = \vec{Q}_{b_L} - \vec{Q}_{b_R} - \vec{Q}_{WS}
\end{equation}
Similarly we define $\vec{c}$ and $\vec{\mu}$ to be:
\begin{eqnarray}
\vec{c} & = & \vec{Q}_{c_L} - \vec{Q}_{t_R} + \vec{Q}_{WS} \\
\vec{\mu} & = & \vec{Q}_{\mu_L} - \vec{Q}_{\mu_R} - \vec{Q}_{WS}
\end{eqnarray}
Note that $\vec{c}$ has been defined using $t_R$ since we have essentially
swapped the right-handed charm and top quarks. Also the charges of the WS
Higgs field
are added rather than subtracted for up-type quarks.
We observe that:
\begin{equation}
\vec{b}+\vec{c}+\vec{\mu}=\vec{0}
\label{b+c+mu=0}
\end{equation}
Now we can express these charges in terms of those of the Higgs fields. We can
define:
\begin{equation}
\vec{b} = a\vec{Q}_W + b\vec{Q}_T + \vec{Q}_X
\end{equation}
where we have chosen the overall sign of the charges on the
Higgs fields $W$ and $T$
so that $a$ and $b$
are not negative. $\vec{Q}_X$ is the total charges of all other
Higgs fields used to
suppress the bottom mass relative to the top mass. We will assume
that $\vec{Q}_X$ cannot be expressed as a linear combination of $\vec{Q}_W$
and $\vec{Q}_T$. Now eqs.~(\ref{mc/mb}) and (\ref{mmu/mb}) require that:
\begin{eqnarray}
\vec{c} & = & \pm (a+1)\vec{Q}_W \pm b\vec{Q}_T \pm \vec{Q}_X \label{cHiggs} \\
\vec{\mu} & = & \pm a\vec{Q}_W \pm (b+1)\vec{Q}_T \pm \vec{Q}_X \label{muHiggs}
\end{eqnarray}
The presence of the $\pm$ signs is due to the fact that we can
use the fields $W^{\dagger}$ and $T^{\dagger}$ as well as $W$ and
$T$ (unlike in supersymmetric models \cite{leurer}-\cite{pokorski})
in Feynman diagrams like fig.~\ref{MbFull}.

So we can rewrite eq.~(\ref{b+c+mu=0}) as:
\begin{equation}
\left( \begin{array}{c} 3a+1 \\ a+1 \\ a-1 \\ -a-1 \end{array}
\right) \vec{Q}_W +
\left( \begin{array}{c} 3b+1 \\ b-1 \\ b+1 \\ -b-1 \end{array}
\right) \vec{Q}_T +
\left( \begin{array}{c} 3 \\ 1 \\ 1 \\ -1 \end{array}
\right) \vec{Q}_X = \vec{0}
\label{QWTX}
\end{equation}
where the 4 coefficients for each term correspond to the 4 combinations of
signs in front of the terms in eqs.~(\ref{cHiggs}) and (\ref{muHiggs}),
giving 64 cases altogether.

All possible choices of coefficient of $\vec{Q}_X$ are non-zero
and, by assumption, $\vec{Q}_X$ is linearly
independent of $\vec{Q}_W$ and $\vec{Q}_T$; so eq.~(\ref{QWTX}) cannot hold.
We must
therefore conclude that there are no Higgs fields other than
$W$ and $T$ used to
suppress the bottom quark mass relative to the top quark mass. So we must set
$\vec{Q}_X=\vec{0}$. We can now use the fact that $a$ and $b$
are not negative, along
with the assumption that $\vec{Q}_T$ is not directly proportional
to $\vec{Q}_W$, to
conclude that eq.~(\ref{QWTX}) requires that:
\begin{equation}
a=b=1
\end{equation}
and that the combination of signs is chosen so that:
\begin{eqnarray}
\vec{b} & = & \vec{Q}_W + \vec{Q}_T \\
\vec{c} & = & -2\vec{Q}_W + \vec{Q}_T \\
\vec{\mu} & = & \vec{Q}_W - 2\vec{Q}_T
\end{eqnarray}

We note that this immediately implies the reasonably good Planck scale
relation:
\begin{equation}
h_b = \; <W><T> \; \approx \frac{1}{65}
\end{equation}

It is now a simple matter to calculate the charges of the
Higgs fields $W$ and $T$.
We have:
\begin{equation}
\vec{Q}_W = \frac{1}{3}(2\vec{b}+\vec{\mu}) =
		\left( 0,-\frac{1}{2},\frac{1}{2},-\frac{4}{3} \right)
\end{equation}
{}From this we can then calculate:
\begin{equation}
\vec{Q}_T = \vec{b} - \vec{Q}_W =
\left( 0,-\frac{1}{6},\frac{1}{6},-\frac{2}{3} \right)
\end{equation}

We notice that the charges of $W$ and $T$ do not cover the 2 dimensional space
of charges $\frac{y_3}{2}$ and $Q_f$, since only even $Q_f$ charges can be
constructed with integer numbers of these Higgs fields.
Therefore, since both $W$
and $T$ have $\frac{y_1}{2}=0$, we will
need at least 2 more Higgs fields to fully cover the 3 dimensional charge space
required to break $G$ down to the SMG. We will now choose 2
more Higgs fields which, together with $W$ and $T$,
will fully cover this space.

Another parameter in the SM, which is within one order of magnitude
from unity, is the mixing matrix element between the 1st and 2nd generations:
\begin{equation}
V_{us} \equiv <\xi> \approx 0.22
\label{Vus}
\end{equation}
With the mass matrix texture in our model, $V_{us}$ is
approximately given by the ratio of the
mass matrix transition element from $d_L$ to
$s_R$ to the transition from $s_L$ to $s_R$. This means that we must have:
\begin{equation}
\vec{Q}_{\xi} = \vec{Q}_{d_L} - \vec{Q}_{s_L}
	= \left( \frac{1}{6},0,0,0 \right) - \left( 0,\frac{1}{6},0,0 \right)
	= \left( \frac{1}{6},-\frac{1}{6},0,0 \right)
\end{equation}

We must now choose one more Higgs field to fully span
the 3 dimensional space of
charges. We shall choose this Higgs field, $\chi$,
so that the transition from $d_L$
to $s_R$ is of the same order of magnitude as the transition from $s_L$ to
$d_R$. This will lead to 2 different but comparable mechanisms for the down
quark mass. So we have:
\begin{equation}
<\chi> = 1
\end{equation}
and the charges of $\chi$ are given by:
\begin{eqnarray}
\vec{Q}_{\chi} & = & [\vec{Q}_{s_L} - \vec{Q}_{d_R}]
		- [\vec{Q}_{d_L} - \vec{Q}_{s_R}] \nonumber \\
 & = & \left[ \left( 0,\frac{1}{6},0,0 \right) -
		\left( -\frac{1}{3},0,0,0 \right) \right] -
	\left[ \left( \frac{1}{6},0,0,0 \right) -
		\left( 0,-\frac{1}{3},0,-1 \right) \right] \nonumber \\
 & = & \left( \frac{1}{6},-\frac{1}{6},0,-1 \right)
\end{eqnarray}

We can now calculate the suppression of all elements in the Yukawa matrices.
However, we must first note that, since we have used 4 Higgs fields, we cannot
uniquely resolve the charge differences between left-handed and right-handed
fermions. There will be some combination of the 4 Higgs field
charges which will result in vanishing charge differences.
We must find the smallest combination of
the 4 Higgs fields which results in a
vanishing set of charges $\vec{Q}=0$. To do this we note that all
fermion $U(1)_f$ charge differences are quantised as integers. However, the 3
Higgs fields $W$, $T$ and $\xi$ can only give integer $U(1)_f$
charge differences which are even.
Therefore we must have at least two $\chi$ Higgs fields
involved in the combination. Then we can find
the unique combination of the other 3 Higgs fields which,
together with the two $\chi$ fields, give
net vanishing charge differences $\vec{Q}=0$.
Thus we find that we can only resolve charges in terms of
the 4 Higgs fields modulo
$2\vec{Q}_{\chi}-2\vec{Q}_{\xi}-9\vec{Q}_{T}+3\vec{Q}_{W}$.
Since this involves such large powers
of $T$, there is usually no ambiguity in selecting the combination
of Higgs fields which suppresses the transition the least.
\section{Mass Matrices}
\label{matrices}
We can now easily calculate the entries in the mass matrices, by expressing the
charge differences between the left-handed and
right-handed fermions in terms of the
charges of the Higgs fields. We define the mass matrices
by considering the mass terms in the SM to be given by:
\begin{equation}
{\cal L}=Q_LM_uU_R+Q_LM_dD_R+L_LM_lE_R+{\rm h.c.}
\end{equation}
The mass matrices can be expressed in terms of Yukawa matrices and the WS Higgs
VEV by:
\begin{equation}
M_f = Y_f \frac{<\phi_{WS}>}{\sqrt{2}}
\end{equation}
This leads to the following order of magnitude Yukawa
matrices, where we have written $W$ instead of $<W>$ etc.
(recalling that we have set $<\chi> = 1$).
\begin{eqnarray}
\label{M_u}
Y_u & = & \left(\begin{array}{ccc} WT^2\xi^2 & WT^2\xi & W^2T\xi \\
				   WT^2\xi^3 & WT^2    & W^2T    \\
				   \xi^3     & 1       & WT \end{array}
	  \right) \\
\label{M_d}
Y_d & = & \left(\begin{array}{ccc} WT^2\xi^2 & WT^2\xi & T^3\xi  \\
				   WT^2\xi   & WT^2    & T^3     \\
				   W^2T^4\xi & W^2T^4  & WT \end{array}
	  \right) \\
\label{M_l}
Y_l & = & \left(\begin{array}{ccc} WT^2\xi^2 & WT^2\xi^3 & WT^4\xi \\
				   WT^2\xi^5 & WT^2    & WT^4\xi^2 \\
				   WT^5\xi^3 & W^2T^4  & WT \end{array}
	  \right)
\end{eqnarray}
\section{Important Features of the Mass Matrices}
\label{features}
One of the most important observations is that the diagonal elements in all 3
mass matrices are the same. This is simply because the $U(1)$
charge differences between the left-handed and right-handed components of a
fermion are the same for all the fermions
within the same proto-generation (e.g.\ $b_L$ and $b_R$ have the same charge
difference as $\tau_L$ and $\tau_R$). This leads to the prediction that the
fermions within each generation should have order of magnitude degenerate
masses (at the fundamental scale), unless some fermion gets its mass from an
off-diagonal term. So we would predict that:
\begin{equation}
m_b \approx m_{\tau}
\label{mb=mtau}
\end{equation}
but $m_t$ will be larger since the dominant term in $M_u$ is off-diagonal.
Similarly we make the prediction:
\begin{equation}
m_s \approx m_{\mu}
\label{ms=mmu}
\end{equation}

For the first generation it is more complicated. At first it appears that all
3 fermions should have the same order of magnitude mass. However,
in diagonalising $Y_d$, there are two contributions of order
$WT^2\xi^2$ (in units of $<\phi_{WS}>/\sqrt{2}$) to the smallest
eigenvalue $m_d$. These come from the element $(Y_d)_{11}$ and
the combination $(Y_d)_{12}(Y_d)_{21}/(Y_d)_{22}$ respectively.
The down quark mass will therefore generally be larger (by
approximately a factor of 2) and we have:
\begin{equation}
m_d \ge m_u \approx m_e
\label{mu=me}
\end{equation}

We can make approximate predictions for the values of all the fermion
masses and also for the values of the mixing angles. However, it is simpler to
calculate everything directly by computer. We will do this by varying the 3
VEVs and finding the best fit to the experimental data.
\section{Results}
\label{results}
Now we are able to choose specific values for the 3 VEVs and calculate the
resulting masses and mixing angles. In order to find the best possible fit we
must use some function which measures how good a fit is. Since we are expecting
an order of magnitude fit, this function should depend only on the ratios of
the fitted masses to the experimentally determined masses. The obvious choice
for such a function is:
\begin{equation}
\chi^2=\sum \left[\ln \left(\frac{m}{m_{\mbox{\small{exp}}}} \right) \right]^2
\end{equation}
where $m$ are the fitted masses and mixing angles and
$m_{\mbox{\small{exp}}}$ are the
corresponding experimental values. The Yukawa
matrices are calculated at the fundamental scale which we take to be the
Planck scale. We use the first order renormalisation group equations (RGEs) for
the SM to calculate the matrices at lower scales (e.g.\ see \cite{RGEs}).
Running masses are calculated in terms of the Yukawa couplings at 1 GeV using
the relation:
\begin{equation}
m(1 {\rm \; GeV})=\frac{h(1 {\rm \; GeV})}{\sqrt{2}}<\phi_{WS}>
\end{equation}
where the low energy VEV of the WS Higgs is:
\begin{equation}
<\phi_{WS}>=246 \; {\rm GeV}
\end{equation}
The only exception is the top quark, where the experimentally measured mass is
the pole mass. For quarks, the pole mass $M$ is related to the
running mass $m$, to first order, by:
\begin{equation}
M=m(M)\left(1+\frac{4}{3}\frac{\alpha_S(M)}{\pi}\right)
\end{equation}

We cannot simply use the 3 matrices given by
eqs.(\ref{M_u})-(\ref{M_l}) to calculate the masses and mixing angles, since
only the order of magnitude of the elements is defined. This could result in
accidental cancellations if we calculated the eigenvalues and eigenvectors
using these values. Therefore we calculate statistically, by giving each
element a random complex phase and then finding the masses and mixing angles.
We repeat this several times and calculate the geometrical mean
for each mass and mixing
angle. In fact we also vary the magnitude of each element randomly by
multiplying by a factor chosen to be the exponential of a number picked from a
Gaussian distribution with mean value 0 and standard deviation 1.

We then vary the 3 free parameters to find the best fit given by the $\chi^2$
function. We get the lowest value of $\chi^2$ for the VEVs:
\begin{equation}
<W> \; = \; 0.158 \; \; <T> \; = \; 0.081 \; \; <{\xi}> \; = \; 0.099
\end{equation}
The fitted value of $<\xi>$ is approximately a factor of two smaller
than the estimate given in eq. (\ref{Vus}). This is mainly because there are
contributions to $V_{us}$ of the same order of magnitude from
both $Y_u$ and $Y_d$.
The result of the fit is shown in table~\ref{bestfit}.
The experimental values
were obtained from \cite{RGEs,PDG}. This fit has a value of:
\begin{equation}
\chi^2=1.68
\end{equation}
This is equivalent to fitting 9 degrees of
freedom (9 masses + 3 mixing angles - 3
Higgs VEVs) to within a factor of
1.54 of the experimental value. This is better than would have been
expected from an order of magnitude fit and should be compared with
$\chi^2=3.7$ for the fit with only 7 degrees of freedom in \cite{Gerry2}.
\begin{table}
\caption{Best fit to experimental data. All masses are running masses at 1 GeV
except the top quark mass which is the pole mass.}
\begin{displaymath}
\begin{array}{|c|c|c|}
\hline
 & {\rm Fitted} & {\rm Experimental} \\ \hline
m_u & 3.8 {\rm \; MeV} & 4 {\rm \; MeV} \\ \hline
m_d & 7.4 {\rm \; MeV} & 9 {\rm \; MeV} \\ \hline
m_e & 1.0 {\rm \; MeV} & 0.5 {\rm \; MeV} \\ \hline
m_c & 0.83 {\rm \; GeV} & 1.4 {\rm \; GeV} \\ \hline
m_s & 415 {\rm \; MeV} & 200 {\rm \; MeV} \\ \hline
m_{\mu} & 103 {\rm \; MeV} & 105 {\rm \; MeV} \\ \hline
M_t & 187 {\rm \; GeV} & 180 {\rm \; GeV} \\ \hline
m_b & 7.6 {\rm \; GeV} & 6.3 {\rm \; GeV} \\ \hline
m_{\tau} & 1.32 {\rm \; GeV} & 1.78 {\rm \; GeV} \\ \hline
V_{us} & 0.18 & 0.22 \\ \hline
V_{cb} & 0.029 & 0.041 \\ \hline
V_{ub} & 0.0030 & 0.002 - 0.005 \\ \hline
\end{array}
\end{displaymath}
\label{bestfit}
\end{table}
\section{Conclusions}
\label{conclusions}
We have presented a model where we can fit the 9 fermion masses and 3 measured
mixing angles using just 3 free parameters, which are the VEVs of 3 fundamental
Higgs fields used to break the model down to the SM.
We would like to conclude by
giving evidence that this fit is significant enough to support such a model.
To do this we shall highlight the important features necessary for our model
to work.

In some ways our main prediction is that all 3 mass matrices have the same
order of magnitudes for corresponding elements on the main diagonal. This leads
to the conclusion that the 3 fermion masses within each generation should be
order of magnitude degenerate. However, the off-diagonal elements are not the
same in the different matrices; so it happens that in the up-type matrix an
off-diagonal element dominates and gives a top quark
mass larger than the bottom quark or
tau lepton masses. This then leads to the charm quark
also getting a mass different from the
strange quark and the muon. Also, to some extent, the down quark
mass is larger (statistically) than the up quark and
electron masses, since it can come from the usual diagonal element or order of
magnitude equal off-diagonal elements. So we can naturally explain the
relations given by eqs.~(\ref{mb=mtau})-(\ref{mu=me}) and that $m_t \gg m_b$.

This is quite different from predictions of theories such as grand
unified $SU(5)$,
where there should be an exact equality between down-type quarks and the
charged leptons in the same generation. We only predict order of magnitude
equality and, in fact, predict that the electron should have the same order of
magnitude mass as the up quark rather than the down quark.

So we would claim that the evidence for this type of model is strong. However,
we cannot really claim that we could only produce such results with this gauge
group. For example we could consider the gauge group
$SU(3) \otimes SU(2) \otimes U(1)^4$ where we define the 4 $U(1)$ groups in the
same way as in our model. This anomaly free $SMG \otimes U(1)^3$ model
would lead to exactly the same results, since we
only used the abelian charges and defined the non-abelian representations in
terms of them. But we would argue that our model was aesthetically better,
since the quantisation of the weak hypercharge is more natural if we have
non-abelian groups $SU(3)_i$ and $SU(2)_i$ associated with
the three $U(1)_i$. The $U(1)_f$ charges are less aesthetically satisfying,
because the $W$ and $T$ Higgs fields have $U(1)_f$ charges which are
quantised in units of one-third of those of the fermions.

\vspace{1cm}
\centerline{\bf Acknowledgements}
\vspace{.3cm}

DJS wishes to acknowledge The Royal Society for funding. HBN
acknowledges funding from INTAS 93-3316, EF contract SC1 0340 (TSTS) and
Cernf{\o}lgeforskning. CF acknowledges funding from
INTAS 93-3316 and PPARC GR/J21231.


\begin{thebibliography}{}
%
\bibitem{fn}
C.\ D.\ Froggatt and H.\ B.\ Nielsen, Nucl.\ Phys.\ {\bf B147}
(1979) 277;
{\bf B164} (1979) 144.
%
\bibitem{bijnens}
J.\ Bijnens and C.\ Wetterich, Nucl.\ Phys.\ {\bf B283} (1987) 237.
%
\bibitem{leurer}
M.\ Leurer, Y.\ Nir and N.\ Seiberg, Nucl.\ Phys.\ {\bf B398} (1993) 319;
{\em ibid} {\bf B420} (1994) 468.
%
\bibitem{ibanezross}
L.\ E.\ Ibanez and G.\ G.\ Ross, Phys.\ Lett.\ {\bf B332} (1994) 100.
%
\bibitem{binetruyramond}
P.\ Bin\'{e}truy and P.\ Ramond, Phys.\ Lett.\ {\bf B350} (1995);
P.\ Bin\'{e}truy, S.\ Lavignac and P.\ Ramond,
preprint LPTHE-ORSAY 95/54, hep-ph/9601243.
%
\bibitem{pokorski}
E.\ Dudas, S.\ Pokorski and C.\ Savoy, Phys.\ Lett.\ {\bf B356} (1995) 45.
%
\bibitem{Bennett}
D.\ L.\ Bennett, H.\ B.\ Nielsen and I.\ Picek,
Phys.\ Lett.\ {\bf B208} (1988) 275;
D.\ L.\ Bennett and H.\ B.\ Nielsen, Int. J. Mod. Phys. {\bf A9}
(1994) 5155;
L.\ V.\ Laperashvili, Yad. Fiz. {\bf 57} (1994) 501
%
%
\bibitem{Gerry2} C.\ D.\ Froggatt, G.\ Lowe and H.B.\ Nielsen, Nucl.\
Phys.\ {\bf B414} (1994) 579.
%
%
\bibitem{corfu} C.\ D.\ Froggatt, to be published in
{\em Proceedings of the Fifth Hellenic School and
Workshops on Elementary Particle Physics},
Corfu, 1995, ed. G.\ Koutsoumbas and N.\ Tracas,
preprint GUTPA/96/02/1, hep-ph/9603432 .
%
\bibitem{RGEs} H. Arason, D.J. Casta\~{n}o, B. Kesthelyi, S. Mikaelian,
E.J. Piard, P. Ramond and B.D. Wright, Phys. Rev. {\bf D46} (1992) 3945
\bibitem{PDG} Particle Data Group, Phys. Rev. {\bf D50} (1994) 1173
%
\end{thebibliography}
\end{document}